\documentclass[
  twocolumn,aps,pra,showpacs,amsmath,
  amssymb,superscriptaddress,bibnotes,
  ]{revtex4-1} 

\usepackage{hyperref}
\usepackage{graphicx}
\usepackage{mathtools}
\usepackage{hyperref}
\usepackage{color}

\usepackage[export]{adjustbox}

\newcommand{\ssa}{{n}}
\newcommand{\ssb}{{m}}

\newcommand{\mbf}[1]{\mathbf{#1}}

\newcommand{\ba}{b^{\phantom{\dagger}}}
\newcommand{\bc}{b^\dagger}

\newcommand{\n}{\hat{n}}

\newcommand{\Tr}{\textrm{Tr}}

\newcommand{\Sp}{\hat{\Sigma}}
\newcommand{\Gp}{\hat{G}}
\newcommand{\Hloc}{\hat{H}}
\newcommand{\astcycl}{\mathrlap{\kern0.085em{\circlearrowright}}\ast}

\newcommand{\Z}{\mathcal{Z}}

\newcommand{\taucycl}{\mathrlap{\kern0.42em{\bullet}}\circlearrowright}

\newcommand{\ttau}{\urcorner}
\newcommand{\taut}{\ulcorner}

\newcommand{\la}{\langle}
\newcommand{\ra}{\rangle}

\begin{document}

\title{Beyond the Hubbard bands in strongly correlated lattice bosons}

\author{Hugo U.~R.~Strand}
\email{hugo.strand@unifr.ch}
\affiliation{Department of Physics, University of Fribourg, 1700 Fribourg, Switzerland} 

\author{Martin Eckstein}
\affiliation{Max Planck Research Department for Structural Dynamics, University of Hamburg-CFEL, 22761 Hamburg, Germany} 

\author{Philipp Werner}
\email{philipp.werner@unifr.ch}
\affiliation{Department of Physics, University of Fribourg, 1700 Fribourg, Switzerland} 

\date{\today} 
\pacs{67.85.-d, 71.10.Fd, 05.30.Jp, 05.50.+q}




\begin{abstract}
We investigate features in the single-particle spectral function beyond the Hubbard bands in the strongly correlated normal phase of the Bose-Hubbard model. There are two distinct classes of additional peaks generated by the bosonic statistics. The first type is thermally activated Hubbard ``sidebands'', with the same physical origin as the zero-temperature Hubbard bands, but generated by excitations from thermally activated local occupation number states. The second class are two-particle fluctuation resonances driven by the lattice dynamics. In the unity filling Mott insulator, this takes the form of a localized triplon combined with a dispersing holon.
Both types of resonances also manifest themselves in the structure factor
and the interaction modulation spectra obtained from nonequilibrium bosonic dynamical mean-field theory calculations.
Our findings explain experimental lattice modulation and Bragg spectroscopy results, and they predict a strong temperature dependence of the first sideband, thereby opening the door to precise thermometry of strongly correlated lattice bosons.
\end{abstract}

\maketitle
\makeatletter
\let\toc@pre\relax
\let\toc@post\relax
\makeatother

\section{Introduction}


Interacting lattice bosons are an active field of research, spurred by continuous experimental advances in cold atom systems \cite{Morsch:2006vn, Bloch:2008uq}. Nowadays, not only are the experimental parameters such as interactions and optical lattice depths well controlled \cite{Jaksch:1998vn}, but there has also been great progress in the field of cold atom spectroscopy \cite{Campbell:2006aa, Stewart:2008aa, Dao:2007aa, Papp:2008aa, Veeravalli:2008aa}, giving access to dynamical quantities such as the structure factor and the single-particle spectral function \cite{Oosten:2005aa, Rey:2005aa, Konabe:2006aa, Fort:2011aa, Fabbri:2012aa}.
In the strongly interacting Mott regime, there are experimental reports on high-energy features in the absorption spectra obtained by lattice modulation \cite{Stoferle:2004aa, Mark:2011aa} and Bragg spectroscopy \cite{Clement:2009aa, Clement:2010aa}. The features appear beyond the first Hubbard resonances at higher multiples of the local interaction energy, and they are generally interpreted as multiple occupation number fluctuations. However, a detailed understanding of the underlying physics is lacking.
%
%


The lowest Hubbard resonances correspond to quasiparticle and quasihole excitations, whose dispersion can be understood already in mean-field and slave-particle approaches \cite{Oosten:2001aa, Dickerscheid:2003aa, Sengupta:2005zr, Huber:2007ys, Menotti:2008tg, Reischl:2005aa}.
The full spectral function and structure factor of the first Hubbard satellites, including finite lifetime broadening, have so far only been settled in one dimension using numerically exact lattice quantum Monte Carlo (QMC) \cite{Batrouni:2005aa, Pippan:2009aa} and density matrix renormalization group (DMRG) calculations \cite{Ejima:2012ab}. However, there are several indications that spectral features beyond the Hubbard bands should exist.
They have been found in the spectral function at zero temperature using the variational cluster approach (VCA) \cite{Knap:2010aa, Knap:2010ab} and strong-coupling calculations \cite{Ejima:2012ab}, and they are also reproduced in the current and kinetic energy susceptibilities calculated by DMRG \cite{Munster:2014aa}.
The recently renewed interest in high-dimensional lattice bosons out-of-equilibrium, to realize complex effective Hamiltonians including
gauge fields \cite{Struck:2012aa,Greschner:2014aa,Goldman:2014aa} and spin-orbit interactions \cite{Lin:2011aa, Struck:2014aa, Jimenez-Garcia:2015aa} by nonequilibrium driving, and the recent real-time generalization \cite{Strand:2015aa} of bosonic dynamical mean-field theory (BDMFT) \cite{Byczuk:2008nx, Hubener:2009cr, Hu:2009qf, Anders:2010uq, Anders:2011uq}, call for a systematic investigation of the positions, origins, and temperature behaviors of these high-energy fluctuations.


In this study, we investigate the nature of the resonances of the spectral function beyond the Hubbard bands at arbitrary integer filling and nonzero temperature.
We also propose an interaction modulation spectroscopy experiment, and we show using nonequilibrium real-time BDMFT \cite{Aoki:2014kx, Strand:2015aa} that it probes the two-particle scattering susceptibility.
Our calculations explain the experimental observations from lattice modulation and Bragg spectroscopy \cite{Stoferle:2004aa, Mark:2011aa, Clement:2009aa, Clement:2010aa}, while making additional predictions regarding the temperature dependence, going beyond previous zero-temperature calculations in one dimension \cite{Kollath:2006nx}.

\section{Model}

We study the strongly correlated limit of cold atoms in the first band of a deep optical lattice, by means of the Bose-Hubbard model \cite{Fisher:1989kl} with Hamiltonian\\[-4mm]
\begin{equation}
  H = - J \sum_{\la i,j \ra} ( \bc_i \ba_j + \bc_j \ba_i )
  + \frac{U}{2} \sum_i \bc_i \bc_i \ba_i \ba_i
  \, ,
\end{equation}\\[-4mm]
where $\bc_i$ ($\ba_i$) creates (annihilates) a boson on site $i$,
$U$ is the local interaction due to two-particle s-wave scattering of the neutral bosonic atoms, and $J$ is the nearest neighbor lattice hopping integral, which we take as our unit of energy.
We further limit the study to the normal phase without symmetry breaking on the Bethe lattice in the limit of infinite dimensions.

In this limit the Bose-Hubbard model is described exactly by BDMFT \cite{Byczuk:2008nx, Hubener:2009cr, Hu:2009qf, Anders:2010uq, Anders:2011uq}, in direct analogy with fermions \cite{Georges:1996aa}.
The model has a direct bearing on cold atoms in three-dimensional cubic lattices, while still being simple enough to allow an understanding of the spectral function by analytic arguments. 
%
%
In BDMFT the lattice is mapped to an interacting impurity coupled to a dynamic external bath with hybridization function $\Delta(\omega)$. For the Bethe lattice, the self-consistency directly relates $\Delta$ to the local Green's function of the impurity $G(\omega)$ according to $\Delta(\omega) = J^2 G(\omega)$ \cite{Georges:1996aa}.

The impurity model can be solved exactly for nonzero temperatures in imaginary time using the continuous-time quantum Monte Carlo method (CT-QMC) \cite{Anders:2010uq}.
However, Monte Carlo approaches are not able to resolve spectral features at high $\omega$ due to the required analytical continuation \cite{Gull:2011lr}, nor are they suitable for out-of-equilibrium calculations due to the dynamical sign problem \cite{Werner:2009tg}.
Instead, we apply a zeroth- and first-order strong-coupling expansion in $\Delta$, namely the Hubbard-I and non-crossing approximation (NCA), respectively \cite{Eckstein:2010fk, Strand:2015aa}, which enables us to directly calculate real-time and real-frequency response functions.
While the NCA is limited to strong interactions, already in this limit there are nontrivial additional single-particle excitations in the Bose-Hubbard model, as we will show.

\section{Method}

The real-time generalization of BDMFT+NCA to nonequilibrium situations has been published elsewhere \cite{Strand:2015aa}.
Here we specialize to the equilibrium case, where the equations can be Fourier-transformed to real frequency; see Appendices \ref{app:RealTime} and \ref{app:RealFreq}.
The NCA amounts to mapping the local impurity Fock space to pseudoparticles and then performing a first-order resummation of hybridization events.
On the real-frequency axis, the single-particle spectral function $A$ is given by the bubble diagram,
\begin{multline}
  A(\omega) = 
  -\frac{1}{(2 \pi)^2} \int_{-\infty}^\infty d\epsilon \,
  \Big( \Tr [ \Gp^<(\epsilon) b \, \Gp^>(\epsilon + \omega) \bc ]
  \\ - \Tr [ \Gp^<(\epsilon) \bc \Gp^>(\epsilon - \omega) b ] \Big)
  \, , \label{eq:G}
\end{multline}
in terms of the pseudoparticle propagator $\Gp$, corresponding to a concomitant fluctuation in occupation number states $n$ and $n \pm 1$ on the impurity.
The pseudoparticle self-energy $\Sp^>$ is given by the shell diagrams with a forward- and backward-propagating hybridization function, and it takes the real-frequency form
\begin{multline}
  \Sp^>(\omega) =
  J^2 \int_{-\infty}^\infty d\epsilon \, \Big(
  f(\epsilon) A(\epsilon)
  [ b \Gp^>(\omega + \epsilon) \bc ] \\ +
  [ 1 + f(\epsilon)] A(\epsilon)
  [ \bc \Gp^>(\omega - \epsilon) b ] \Big)
  \label{eq:Sigma} \, ,
\end{multline}
where $f(\omega) = (e^{\beta \omega} - 1)^{-1}$ is the Bose distribution function with inverse temperature $\beta$, and the BDMFT lattice self-consistency has been used; for details see Appendix \ref{app:RealFreq}. 
The first term in Eq.~(\ref{eq:Sigma}) describes a particle fluctuation on the lattice involving the occupied density of states $f(\omega) A(\omega)$, while the second term is a hole fluctuation with the unoccupied density of states $[1 + f(\omega)]  A(\omega)$.

\section{Results}

\subsection{Analytcial considerations}

\subsubsection{Hubbard bands}


The main spectral features of the Bose-Hubbard model can be understood already at zeroth order in $\Delta$.
This amounts to the Hubbard-I approximation (HIA) \cite{Hubbard:1963aa}
where the lattice self-energy $\Sigma(\omega, \mbf{k})$ is approximated with the zero-hopping ($J=0$) self-energy $\Sigma_{\textrm{HIA}}(\omega)$, $\Sigma(\omega, \mbf{k}) \approx \Sigma_{\textrm{HIA}}(\omega)$.
The self-energy can be determined analytically, see Appendix \ref{app:HIA}, and it takes the form
\begin{equation}
  \Sigma_{\textrm{HIA}}(z) = [2Un(z + \mu) - U^2n(n-1)]/(z + \mu + U)
  \, ,
\end{equation}
at zero temperature, with a single pole at negative frequencies. 
Reinsertion into the lattice Green's function $G(z, \mbf{k}) \approx [z + \mu - \epsilon_{\mbf{k}} - \Sigma_{\textrm{HIA}}(z)]^{-1}$, with the noninteracting single-particle dispersion $\epsilon_\mbf{k}$, produces two excitation branches with dispersion,
\begin{multline}
  2\tilde{\epsilon}_{\mbf{k}}
  = \epsilon_{\mbf{k}} + U(2n - 1) - 2\mu
  \\
    \pm \sqrt{ \epsilon_{\mbf{k}}^2 + 2U(2n + 1) \epsilon_{\mbf{k}} + U^2 }
  \, .
\end{multline}
Expanding to third order in $J/U$ the bandwidths $\tilde{W}$ of the two branches are given in terms of the noninteracting bandwidth $W$ as $\tilde{W} = W(n+1)$ and $Wn$. The upper and lower branches are centered at $\bar{\epsilon} = U(2n - 1 \pm 1)/2 - \mu$ respectively. These two spectral features are the bosonic analog of the Hubbard bands in the fermionic Hubbard model corresponding to the local particle and hole excitations $|n \ra \rightarrow | n \pm 1 \ra$. 
We note that the simple Hubbard-I approximation reproduces previous mean-field and slave-particle results \cite{Oosten:2001aa, Dickerscheid:2003aa, Sengupta:2005zr, Huber:2007ys, Menotti:2008tg, Reischl:2005aa} in the normal phase.


\begin{figure}
\includegraphics[scale=1.0] {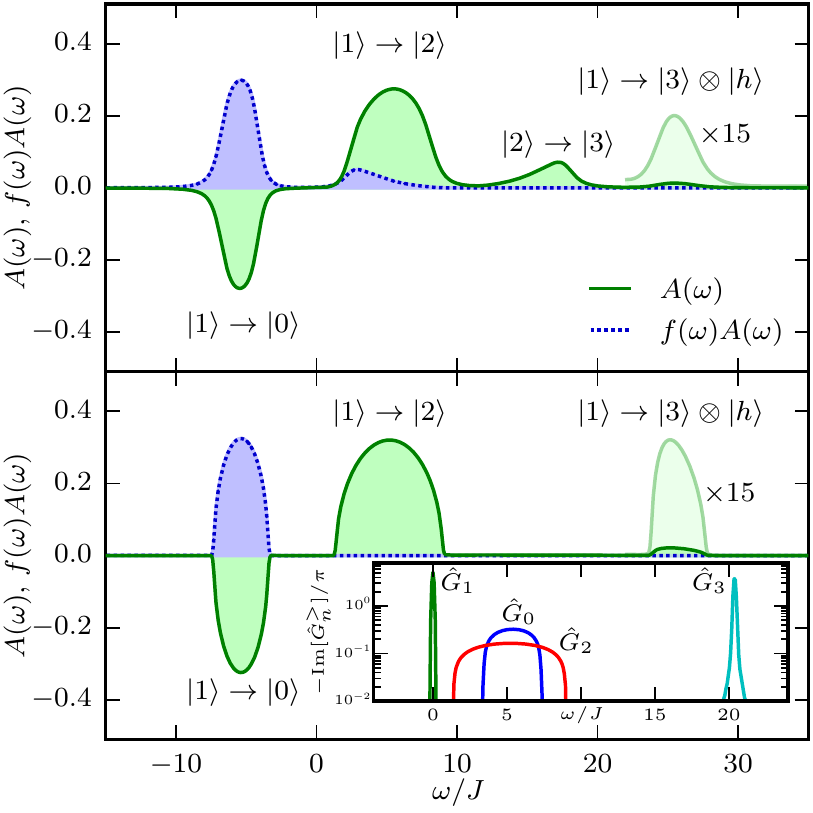} \\
\caption{\label{fig:n1} (Color online) Bose-Hubbard model single-particle spectral function $A(\omega)$ (green solid line) for $J=1$, $U=10$, $\mu=U/2$ and the occupied density of states $f(\omega) A(\omega)$ (blue dotted line), at high temperature $T=2$ (upper panel) and low temperature $T=1/4$ (lower panel). Inset: pseudoparticle Green's functions $-\textrm{Im}[\Gp_n^>(\omega)]/\pi$ for the lowest occupation number states $n = 0$ -- $3$ at $T=1/4$.
} \end{figure}

In the fermionic Hubbard model, the effect of dynamical lattice fluctuations is to modify the shape of the Hubbard bands. In the Bose-Hubbard model, however, the bosonic statistics does not restrict the occupation of a single state, and additional fluctuations become relevant.
This is directly evident when studying the unity filling $n=1$ spectral function using BDMFT+NCA; see the upper panel in Fig.\ \ref{fig:n1}. The upper and lower Hubbard bands are located at $\omega \approx \pm U/2$ with approximate bandwidths $\tilde{W} \approx 2W$ and $W$, respectively, with the noninteracting bandwidth being $W = 4J$, in accordance with the Hubbard-I approximation.

\subsubsection{Additional spectral features}

Interestingly, in addition to the Hubbard bands, there are two additional spectral features above the upper Hubbard band at $\omega \approx 3U/2$ and $5U/2$.
%
%
The origin of these two spectral features can, in part, be understood by taking the zero-temperature limit of the NCA diagrams in Eqs.\ (\ref{eq:G}) and (\ref{eq:Sigma}). In this limit, the lesser pseudoparticle Green's function reduces to the integer-filling $n$ Fock state,
\begin{equation}
  \Gp^<(\omega) \approx -i 2\pi | n \ra \delta(\omega) \la n |
  \, ,
\end{equation}
and insertion into Eq.\ (\ref{eq:G}) gives the spectral function as
\begin{equation}
  -i2\pi A(\omega) \approx (n+1) \Gp^>_{n+1}(\omega) - n \Gp^>_{n-1}(-\omega)
  \, ,
\end{equation}
where $\Gp^>_n$ is the $n$th occupation number state propagator. At unity filling $n=1$ this simple structure of the spectral function shows that the spectral weight at negative frequencies is due to the holon propagator $\Gp_0$, while at positive frequencies it is due to the doublon propagator $\Gp_2$.
%
%
Also the pseudoparticle self-energy $\Sp^>$ in Eq.\ (\ref{eq:Sigma}) simplifies by noting that the occupied and unoccupied density of states becomes
\begin{align}
  -i2\pi f(\omega) A(\omega) & \approx (n+1)\Gp^>_{n+1}(\omega)
  \, , \\
  -i2\pi [1 + f(\omega)] A(\omega) & \approx -n\Gp^>_{n-1}(-\omega)
  \, .
\end{align}
Insertion into Eq.\ (\ref{eq:Sigma}) gives
\begin{multline}
  \Sp^>_m(\omega) \approx
  \frac{i J^2}{2\pi} \int_{-\infty}^\infty d\epsilon \,
  \Big[ n (m+1) \Gp^>_{n-1}(\omega - \epsilon) \Gp^>_{m+1}(\epsilon) \\ +
  (n+1) m \, \Gp^>_{n+1}(\omega - \epsilon) \Gp^>_{m-1}(\epsilon) \Big]  
  \, . \label{eq:SigmaGtrM}
\end{multline}
Hence, at unity filling the holon and doublon self-energies become
\begin{align}
  \Sp^>_0(\omega) & \approx J^2 \Gp^>_0(\omega) \, , \label{eq:SigmaGtr0} \\
  \Sp^>_2(\omega) & \approx 4 J^2 \Gp^>_2(\omega) + J^2 3 \Gp^>_0(\omega - 2U) \, ,
  \label{eq:SigmaGtr2}
\end{align}
where in the last relation we have used that the local triplon pseudoparticle $\Gp^>_3$ is long-lived, i.e., $\Gp^>_3(\omega) = 1/(\omega + i\eta - 2U)$, as confirmed by numerical calculations; see inset in Fig.\ \ref{fig:n1} (and Appendix \ref{app:Anal} for the generalization to arbitrary filling $n$).
We see that the holon self-energy $\Sp^>_0$ only depends self-consistently on the holon Green's function $\Gp^>_0$. In analogy to the lattice self-consistency for the Bethe lattice, this results in a semicircular holon pseudoparticle spectral function. The doublon self-energy $\Sp^>_2$ has a similar dependence on $\Gp^>_2$ but with an additional triplon term containing the holon $\Gp^>_0$ with an effective frequency shift of the triplon plus holon energy $2U + U/2 = 5U/2$.

\subsection{Bosonic dynamical mean field theory}

The full BDMFT+NCA calculation at low temperature confirms these analytic arguments; see the lower panel of Fig.\ \ref{fig:n1}. The lower Hubbard band is produced by the hole excitation $|1\ra \rightarrow |0\ra$ described by the holon propagator $-\Gp_0(-\omega)$ with a semicircular shape and bandwidth $4J$. The positive frequency spectrum is given by the doublon propagator $2\Gp_2(\omega)$ with an almost semicircular upper Hubbard band ($|1\ra \rightarrow |2\ra$) and a local triplon excitation combined with a delocalized holon ($|1 \ra \rightarrow |3\ra \otimes |h\ra$).
We also note that the high-temperature peak at $\omega \approx 3U/2$ is not present at low temperature; see the lower panel of Fig.\ \ref{fig:n1}.

At nonzero temperature, the local occupation on the lattice fluctuates around the mean filling, and the simplifying approximation of the occupied states in the pseudoparticle propagator breaks down as both the holon and doublon will contribute. However, the doublon is easier to activate thermally due to its two times larger bandwidth compared to the holon. The significant doublon contribution is seen in the occupied density of states of the upper Hubbard band in the upper panel of Fig.\ \ref{fig:n1}. The doublons also contribute to the spectral function, which corresponds to a $\pm 1$ particle excitation on the ground state. Hence, the thermally activated doublons undergo the excitations $|2\ra \rightarrow |2+(\pm 1)\ra$, where the hole excitation ($|2\ra \rightarrow |1\ra$) modifies the lower Hubbard band and the particle excitation ($|2\ra \rightarrow |3\ra$) creates a long-lived triplon (without a holon this time) yielding the new spectral feature at $\omega \approx 3U/2$. The inverted shape of the $|2\ra \rightarrow |3\ra$ peak as compared to the thermal occupation of the upper Hubbard band can also be understood as the doublons in the ground state are being excited to the triplon with the excitation energy $U - \tilde{\epsilon}_k$, hence the low-energy doublons contribute to the high-energy part of the peak.

\begin{figure}
\includegraphics[scale=1.00] {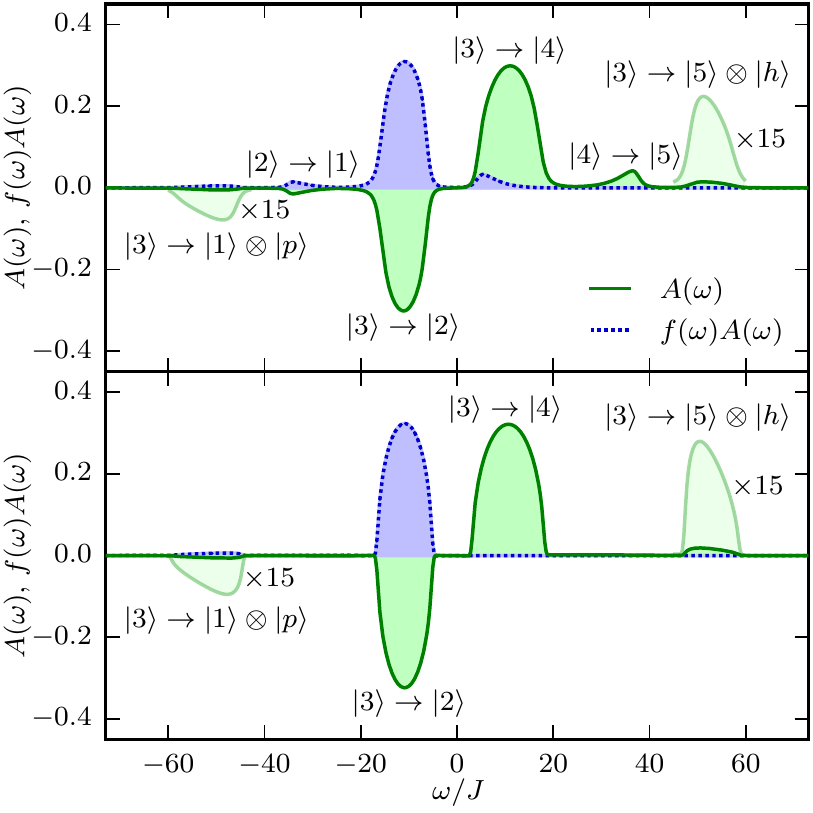} \\
\caption{\label{fig:n3}(Color online) Bose-Hubbard model real-frequency single-particle spectral function $A(\omega)$ (green solid line) and occupied density of states $f(\omega) A(\omega)$ (blue dotted line) at three boson filling $n=3$, $J=1$, $U=20$, and $\mu=5U/2$ at high temperature $T=3$ (upper panel), and low temperature $T=1/4$ (lower panel).
} \end{figure}

The additional spectral features described here, namely (i) the zero-temperature double-particle excitation $| n \ra \rightarrow | n+2 \ra \otimes |(n-1)_{disp}\ra$ (triplon with dispersing hole at unity filling) and (ii) thermally activated side bands $| n+1 \ra \rightarrow | n+2 \ra$, are fundamental contributions to the Bose-Hubbard spectral function and generalize to any integer filling. Unity filling is a special case which is nonsymmetric in particle and hole excitations. At higher fillings also a double hole excitation is present, $|n\ra \rightarrow |n-2\ra \otimes |(n+1)_{disp}\ra$, and thermally activated sidebands proliferate in both positive and negative frequency as temperature is increased, $| n \pm m \ra \rightarrow | n \pm (m + 1) \ra, \, | n \pm (m - 1) \ra$ (with $m \le n$); see Fig.\ \ref{fig:n3} for the case of $n=3$.

\subsection{Modulation spectroscopy}

\begin{figure}
  \includegraphics[scale=1] {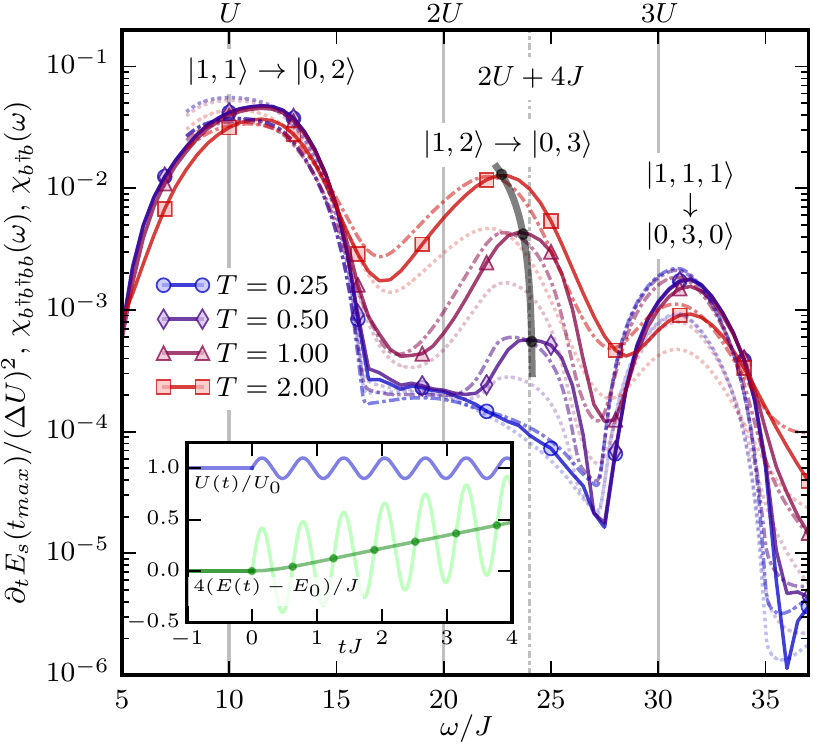} \\
  \caption{\label{fig:absexp}(Color online) Interaction modulation energy absorption spectra (solid lines) compared to the local two-particle susceptibility $\chi_{\bc\bc bb}(\omega)$ (dash-dotted lines) and density susceptibility $\chi_{\bc b}(\omega)$ (dotted lines) for the temperatures $T=0.25$, $0.5$, $1$, and $2$ (circles, diamonds, triangles, and squares, respectively). Inset: real-time evolution of the interaction $U(t)$ and total energy $E(t)$ for driving frequency $\omega=10$ and $T = 2$. 
} \end{figure}

To connect the results for the spectral function to the particle number conserving system response, we now perform interaction modulation spectroscopy and compute interaction and density susceptibilities.
Using nonequilibrium BDMFT, we calculate the energy time-evolution $E(t)$ while modulating the interaction sinusoidally, $U(t) = U_0 + \Delta U \sin(\omega t)$, with frequency $\omega$ and amplitude $\Delta U / U_0 = 0.1$, during four inverse hoppings $t_{max} = 4/J$.
The energy absorption rate $\partial_t E_{s}(t_{max})$ is determined from interpolated stroboscopical energy measurements $E_s(t)$, see inset of Fig.\ \ref{fig:absexp}.
Sweeping $\omega$ in the range $U/2 < \omega < 7U/2$ at different initial temperatures $T$ produces the spectra of Fig.\ \ref{fig:absexp}, displaying three resonances directly connected to the spectral function features.

The Hubbard-type particle-hole fluctuation ($|1,1 \ra \rightarrow |0,2\ra$) is centered at $\omega \approx U$. The doublon to holon-triplon excitation ($|1,2\ra \rightarrow |0,3\ra$) located at $\omega \approx 2U + 4J$ is strongly thermally activated and features the same inverted distribution as the Hubbard sidebands in Figs.\ \ref{fig:n1} and \ref{fig:n3}. The corresponding peak exhibits the additional shift $4J$, i.e., half the bandwidth of the upper Hubbard band at low $T$ which is reduced to $\omega \approx 2U$ at higher $T$.
The three singlons to two holons and one triplon fluctuation ($|1, 1, 1\ra \rightarrow |0,0,3\ra$) is located at $\omega \approx 3U$. It exhibits a weak $T$ dependence, and is the analog of the triplon plus dispersing holon resonance in the spectral function.
In linear response, $\partial_t E(t)$ is directly related to the interaction susceptibility according to $\partial_t E_s(t) \propto (\Delta U)^2 \chi_{\bc \bc b b}(\omega)$.
We compute the impurity $\chi_{\bc \bc b b}(\omega)$ using equilibrium real-time BDMFT+NCA, and we find quantitative agreement with the absorption spectra apart from a slight red shift; see Fig.\ \ref{fig:absexp}.
To connect to the structure factor, we also compute the density susceptibility $\chi_{\bc b}(\omega)$, which has the same structure as $\chi_{\bc \bc b b}(\omega)$ apart from a $1/2$ reduction of the $2U$ and $3U$ resonances; see Fig.\ \ref{fig:absexp}.

\section{Discussion}

Experimentally, the $U$ and $2U$ resonances have been observed in one dimension using Bragg spectroscopy \cite{Clement:2009aa, Clement:2010aa} and in one and three dimensions using lattice modulation spectroscopy \cite{Stoferle:2004aa, Mark:2011aa}, where also a weak $3U$ resonance has been reported \cite{Stoferle:2004aa}.
In contrast to one-dimensional calculations at zero temperature \cite{Kollath:2006nx}, we find that in higher dimensions and at finite temperatures both $2U$ and $3U$ resonances are present even at commensurate filling.
We also explain the peak shape of the $2U$ resonance and its strong thermal activation, almost two orders of magnitude in Fig.\ \ref{fig:absexp}, confirming indications from restricted basis calculations \cite{Reischl:2005aa}.
Additional experimental work in this direction should be worthwhile, especially in exploring and using the $2U$ resonance for Mott state thermometry. The $3U$ triplon excitation is interesting in its own right and in connection with Efimov physics \cite{Efimov:1970aa,Bedaque:2000aa,Kraemer:2006aa}.
Our findings are also relevant in the context of driven experiments, e.g., for generating gauge fields \cite{Struck:2012aa,Greschner:2014aa,Goldman:2014aa} and spin-orbit interactions \cite{Lin:2011aa, Struck:2014aa, Jimenez-Garcia:2015aa}, where energy absorption at the driving frequency needs to be minimized.
It would also be interesting to study the fine-structure changes in the spectral function induced by the higher bands of the optical lattice through effective multibody interactions \cite{Johnson:2009kx, Will:2010uq, Tiesinga:2011ve} and renormalized hopping \cite{0256-307X-29-8-083701, Zhu:2015aa}.

Theoretically, spectral features beyond the Hubbard bands in the normal phase have been reported in one dimension at zero temperature using VCA, DMRG, and hopping perturbation theory \cite{Knap:2010aa, Knap:2010ab, Ejima:2012ab, Munster:2014aa} without connecting to experiments, while equilibrium temperature effects have been considered on the slave-particle mean-field level \cite{Dickerscheid:2003aa, Reischl:2005aa}.
Within BDMFT in combination with CT-QMC, which is exact for the normal phase in infinite dimensions, one cannot expect to observe these features because of the limitation of analytical continuation \cite{Panas:2015ab}. 

\section{Conclusions}

In summary, we have analyzed the nature of the resonances of the single-particle spectral function of the Bose-Hubbard model in the symmetric Mott phase. We identified two classes of fundamental excitations beyond the Hubbard bands, namely the double-particle excitation (triplon plus dispersing holon at unity filling) and thermally activated sidebands generated by excitations from thermally occupied number states.
These resonances explain the features in interaction modulation spectra and the structure factor, and they are therefore important for the interpretation of experiments \cite{Stoferle:2004aa, Clement:2009aa, Clement:2010aa, Mark:2011aa}.
How these features evolve at weaker interactions and in the symmetry-broken phase is an interesting question for future investigations.

\begin{acknowledgments}
The authors would like to acknowledge fruitful discussions with 
L. V. Boehnke, D. Golez, A. Herrmann, D. H\"{u}gel, Y. Murakami, J. Panas, and L. Pollet.
The calculations have been performed on the UniFr cluster. H.S. and P.W. are supported by the European Research Council under the European Union's Seventh Framework Programme (FP7/2007-2013) / ERC Grant Agreement No. 278023.
%
\end{acknowledgments}

\appendix
\section{Hubbard-I approximation}
\label{app:HIA}

For the Bose-Hubbard model, mean-field calculations in combination with the random-phase approximation (MF+RPA) and slave-particle approaches already produce nontrivial spectral functions \cite{Oosten:2001aa, Dickerscheid:2003aa, Huber:2007ys, Menotti:2008tg}.
In this appendix we rederive these results from a DMFT perspective and show that in the normal phase the spectral functions obtained via these methods are equivalent to the Hubbard-I approximation (HIA) \cite{Hubbard:1963aa}, where the lattice self-energy is approximated with the self-energy in the zero hopping limit $J=0$.
Without hopping, the Bose-Hubbard model separates into a collection of local Hamiltonians $\hat{H} = \frac{1}{2}U\n(\n-1) - \mu\n$. The real-frequency spectral function can then readily be obtained from the Lehmann expansion of the local single-particle Green's function,
\begin{multline}
  G_L(\omega)
  =
  \frac{1}{\Z} \sum_{\ssa \ssb}
  \frac{
  \langle \ssa | b | \ssb \rangle
  \langle \ssb | \bc | \ssa \rangle
  }{\omega + i\eta + E_\ssa - E_\ssb} ( e^{-\beta E_\ssa} - e^{-\beta E_\ssb} )
  \, ,
  \label{eq:Lehmann}
\end{multline}
where $|n\ra$ is the occupation number state with energy $E_n =  \frac{1}{2}Un(n-1) - \mu n$. The Green's function can be decomposed into hole and particle excitation branches according to
\begin{equation*}
  G_L (\omega) =
  G_L^{(p)}(\omega) +
  G_L^{(h)}(\omega)
  \, ,
\end{equation*}
where
\begin{align}
  G_L^{(p)}(\omega) & =
  + \frac{1}{\Z} \sum_{\ssa \ssb} e^{-\beta E_\ssa}
  \frac{
  \langle \ssa | b | \ssb \rangle
  \langle \ssb | \bc | \ssa \rangle
  }{\omega + i\eta + E_\ssa - E_\ssb}
  \, , \\
  G_L^{(h)}(\omega) & =
  - \frac{1}{\Z} \sum_{\ssa \ssb} e^{-\beta E_\ssa}
  \frac{
  \langle \ssa | \bc | \ssb \rangle
  \langle \ssb | b | \ssa \rangle
  }{\omega + i\eta - E_\ssa + E_\ssb}
  \, .
\end{align}
We immediately see that the negative frequency hole excitations $G^{(h)}_L$ come with negative spectral weight for bosons.


For integer filling $n$ and zero temperature Eq.\ (\ref{eq:Lehmann}) simplifies to
\begin{equation}
  G_L(\omega) =
  \frac{n+1}{\omega + i\eta - Un + \mu}
  - \frac{n}{\omega + i\eta - Un + U + \mu}
  \, .
\end{equation}
This gives the Hubbard-I approximation of the lattice self-energy if we use the noninteracting atomic Green's function $G_0(z) = 1/(z + \mu)$ and the Dyson equation $\Sigma(z) = G_0^{-1}(z) - G^{-1}(z)$. The analytic form for the self-energy $\Sigma$ is
\begin{equation}
  \Sigma(z) = \frac{2Un (z+\mu) - U^2n(n - 1)}{z + \mu + U}
  \, . \label{eq:AppSHI}
\end{equation}
Note that the self-energy has a very simple structure with a single pole at $z = -U - \mu$ (e.g., for $n=1$ and $\mu = U/2$ it is located at $z = -3U/2$).

The lattice Green's function $G(\omega, \mathbf{k})$ in the Hubbard-I approximation is now obtained as
\begin{equation}
  G(\omega, \mathbf{k}) = \frac{1}{\omega + i\eta + \mu - \epsilon_\mathbf{k} - \Sigma(\omega)}
  \, ,
\end{equation}
where, as $\Sigma$ is momentum-independent, $G(\omega, \mathbf{k})$ depends only on $\mathbf{k}$ through the noninteracting dispersion $\epsilon_{\mathbf{k}}$. Insertion of the analytic form [Eq.\ (\ref{eq:AppSHI})] generates two pole branches in $G(\omega, \mathbf{k})$ with dispersions
\begin{multline}
  2\tilde{\epsilon}_{\mathbf{k}} = \epsilon_{\mathbf{k}} + U(2n - 1) - 2\mu \\
  \pm \sqrt{ \epsilon_{\mathbf{k}}^2 + 2U(2n + 1) \epsilon_{\mathbf{k}} + U^2 }
  \, ,
\end{multline}
corresponding to the $n \pm 1$ particle and hole excitations, i.e., the upper and lower Hubbard bands, with band centers at $\bar{\epsilon} = U(2n-1) \pm 1)/2 - \mu$.
The bandwidths $\tilde{W}$ of these bands as a function of noninteracting bandwidth $W$ of $\epsilon_{\mathbf{k}} \in \{-W/2, W/2\}$, $\tilde{W} = \tilde{\epsilon}^{(\pm)}_{\mathbf{k}}(W/2) \pm \tilde{\epsilon}_{\mathbf{k}}^{(\pm)}(-W/2)$, is to second order in $W/U$ given by $\tilde{W} = W(n+1)$ and $\tilde{W} = Wn$ for the upper and lower Hubbard band, respectively. Hence the widths scale with the integer filling $n$ and exhibit the largest asymmetry at unity filling $n=1$.
The Hubbard-I approximation directly reproduces the results for the normal phase obtained by MF+RPA and from slave-particle theory \cite{Oosten:2001aa, Dickerscheid:2003aa, Huber:2007ys, Menotti:2008tg}, and it also generalizes trivially to nonzero temperatures.

\section{BDMFT+NCA equilibrium real-time propagation}
\label{app:RealTime}

To calculate real-frequency properties in BDMFT+NCA we adapt the real-time out-of-equilibrium formulation of Ref.\ \onlinecite{Strand:2015aa} to equilibrium, where all response functions are time translation invariant, $G(t, t') = G(t - t')$.
Using the notation of Ref.\ \cite{Aoki:2014kx}, this yields a simplified set of equations for the pseudoparticle real-time propagation.

While it is indeed possible to transform the resulting equations directly to real frequency it turns out to be numerically easier to obtain high-quality real-frequency results by performing a real-time evolution and then Fourier transforming the results to real frequency.
Using this scheme, one avoids inverting the Dyson equation on the real-frequency axis, which requires a careful discretization and handling of Lorentzian broadening factors. (In the low-temperature limit this inversion is increasingly difficult as the pseudoparticle ground state approaches a delta function at zero frequency in the Bose-Hubbard model.)

For the time evolution in equilibrium, only the Dyson equations for the retarded component $\Gp^R(t)$ and the right-mixing component $\Gp^\ttau(t,\tau)$ of the pseudoparticle Green's function $\Gp$ are required, assuming that the Matsubara imaginary-time Green's function $\Gp^M(\tau)$ is known. The Dyson equation for $\Gp^R(t)$ takes the Volterra form
\begin{equation}
  ( i \partial_t - \Hloc ) \Gp^R(t)
  - \int_0^t d\bar{t} \,
  \Sp^R(t - \bar{t}) \Gp^R(\bar{t}) = 0
  \, , \label{eq:AppDR}
\end{equation}
with the initial boundary condition $\Gp^R(0) = -i \mbf{1}$ and the pseudoparticle self-energy $\Sp^R$.
The Dyson equation for $\Gp^\ttau(t, \tau)$ also depends on $\Sp^R$ according to
\begin{equation}
  (i \partial_t  - \Hloc) \Gp^\ttau(t, \tau)
  - \int_0^t d\bar{t} \,
  \Sp^R(t - \bar{t}) \Gp^\ttau(\bar{t}, \tau)
  = Q^\ttau(t, \tau)
  \, , \label{eq:AppDttau}
\end{equation}
where the additional right-hand side $Q^\ttau(t, \tau)$ is given by the Volterra-type convolution of the right-mixing pseudoparticle self-energy $\Sp^\ttau(t, \tau)$ and the imaginary-time pseudoparticle Green's function $\Gp^M(\tau)$:
\begin{equation}
  Q^\ttau(t, \tau) =
  \int_0^\tau d\bar{\tau} \,
  \Sp^\ttau(t,\bar{\tau}) \Gp^M(\bar{\tau})
  \, . \label{eq:AppQttau}
\end{equation}
The lesser Green's function $\Gp^<(t)$ is directly given by the right-mixing Green's function at $\tau=0$, $\Gp^<(t) = \Gp^\ttau(t, 0)$. Furthermore, the projection onto the physical space \cite{Eckstein:2010fk} also gives the (pseudoparticle-specific) relation for the greater component $\Gp^R(t) = \theta(t) \Gp^>(t)$.
Both lesser and greater components  $\Gp^\gtrless(t)$ are readily extended to all times using the antihermicity relation $\Gp^\gtrless(-t) = -[\Gp^\gtrless(t)]^\dagger$ \cite{Aoki:2014kx}. Hence solving Eqs.\ (\ref{eq:AppDR}) and (\ref{eq:AppDttau}) determines all Keldysh components of $\Gp$.

Static local observables are obtained as direct traces over the lesser pseudoparticle Green's function,
\begin{equation}
  \langle \hat{O} \rangle = i \Tr[ \hat{O} \Gp^<(0) ]
  \, ,
\end{equation}
at relative time $t=0$, trivially giving the same result as the imaginary-time equilibrium calculation $\langle \hat{O} \rangle = - \Tr [ \hat{O} \Gp^M(\beta) ] = i \Tr[ \hat{O} \Gp^\ttau(0, 0) ] = i \Tr[ \hat{O} \Gp^<(0) ]$.

In NCA, the lesser and greater single-particle Green's functions $G^\gtrless$ are obtained as the pseudoparticle bubble
\begin{equation}
  G^\gtrless(t) = i \Tr[ \Gp^\lessgtr(-t) b \, \Gp^\gtrless(t) \bc ]
  \, . \label{eq:AppGsp}
\end{equation}
Note that we only consider the normal phase here without symmetry breaking, $\la b \ra = 0$, hence the full Green's function is equal to the connected Green's function.
The form for the single-particle Green's function can be generalized to arbitrary susceptibilities $\chi_{\hat{A}, \hat{B}}(t)$ between a pair of operators $\hat{A}$ and $\hat{B}$ yielding
\begin{multline}
\chi_{\hat{A}, \hat{B}}(t) =
  i \theta(t) \Big(
  \Tr[ \Gp^<(-t) \hat{A} \, \Gp^>(t) \hat{B} ]
  \\ -
  \Tr[ \Gp^>(-t) \hat{B} \, \Gp^<(t) \hat{A} ]
  \Big)
  \, ,
\end{multline}
i.e., $\chi_{b, \bc}(t)$ corresponds to the retarded single-particle Green's function, $\chi_{b, \bc}(t) = G^R(t)$. Note that the equal operator susceptibility is denoted as $\chi_{\hat{A}} \equiv \chi_{\hat{A}, \hat{A}}$.

The retarded pseudoparticle self-energy is obtained from the greater component $\Sp^R(1) = \theta(1) \Sp^>(1)$ which in turn is given by the two shell diagrams,
\begin{multline}
  \Sp^>(t) = i \Big(
  \Delta^>(t) [ \bc \Gp^>(t) b ] \\ +
  \Delta^<(-t) [ b \Gp^>(t) \bc ] \Big)
  \, , \label{eq:AppSgtr}
\end{multline}
and the right-mixing component is obtained analogously,
\begin{multline}
  \Sp^\ttau(t, \tau) = i \Big(
  \Delta^\ttau(t, \tau) [ \bc \Gp^\ttau(t, \tau) b ] \\ +
  \Delta^\taut(\tau, t) [ b \Gp^\ttau(t, \tau) \bc ] \Big)
  \, . \label{eq:AppSttau}
\end{multline}
Use of the Bethe lattice self-consistency relation $\Delta(t) = J^2 G(t)$, then yields the closed set of BDMFT+NCA equations in combination with Eqs.\ (\ref{eq:AppDR}), (\ref{eq:AppDttau}), (\ref{eq:AppQttau}), (\ref{eq:AppGsp}), (\ref{eq:AppSgtr}), and (\ref{eq:AppSttau}).

The equilibrium real-time Dyson equations are solved using an equidistant time-discretized second-order propagation method. While the convolutions scale quadratically with the time step $\mathcal{O}(N_t^2)$ as in the out-of-equilibrium method, the time-translation invariance reduces the memory scaling from quadratic to linear $\mathcal{O}(N_t)$, enabling much longer time calculations, which yield a very fine real-frequency resolution after Fourier transformation.

\section{Real frequency}
\label{app:RealFreq}

While we perform the numerical calculations in real time it turns out that formulating NCA in real frequency is a fruitful venue for understanding the physics, especially the low temperature limit.
Hence we Fourier-transform the equilibrium real-time equations using
\begin{equation*}
  G(\omega) = \int_{-\infty}^\infty  \!\!\!  dt \, e^{i\omega t} G(t)
    \, , \quad
    G(t) = \frac{1}{2\pi} \int_{-\infty}^\infty \!\!\! d\omega \, e^{-i\omega t} G(\omega)
    \, .
\end{equation*}
The single-particle spectral function is obtained from Eq.\ (\ref{eq:AppGsp}) as,
\begin{multline}
  A(\omega) = -\frac{1}{\pi} \textrm{Im} [ G^R(\omega) ] =
  \frac{i}{2 \pi} [ G^>(\omega) - G^<(\omega) ] \\ =
  -\frac{1}{(2 \pi)^2} \int_{-\infty}^\infty d\epsilon \,
  \Big( \Tr [ \Gp^<(\epsilon) b \, \Gp^>(\epsilon + \omega) \bc ]
  \\ - \Tr [ \Gp^<(\epsilon) \bc \Gp^>(\epsilon - \omega) b ] \Big)
  \, , \label{eq:AppbA}
\end{multline}
and the greater pseudoparticle self-energy $\Sp^>$ in Eq.\ (\ref{eq:AppSgtr}) takes the form
\begin{multline}
  \Sp^>(\omega) = \frac{i}{2\pi} \int_{-\infty}^\infty d\epsilon \,
  \Big( \Delta^<(\epsilon) [ b \Gp^>(\omega + \epsilon) \bc ] \\ +
  \Delta^>(\epsilon) [\bc \Gp^>(\omega - \epsilon) b ] \Big)
  \, . \label{eq:AppSigma}
\end{multline}
Now, using the Bethe lattice self-consistency
$\Delta(\omega) = J^2 G(\omega)$ in combination with the relation of the greater and lesser components to the single-particle spectral function, 
$G^>(\omega) = -i 2\pi A(\omega) f(\omega)$ and $G^<(\omega) = -i 2 \pi A(\omega) [ 1 + f(\omega)]$, where $f(\omega) = (e^{\beta \omega} - 1)^{-1}$ is the Bose distribution function, this can be rewritten as
\begin{multline}
  \Sp^>(\omega) =
  J^2
  \int_{-\infty}^\infty d\epsilon \, \Big(
  f(\epsilon) A(\epsilon)
  [ b \Gp^>(\omega + \epsilon) \bc ] \\ +
  [ 1 + f(\epsilon)] A(\epsilon)
  [ \bc \Gp^>(\omega - \epsilon) b ] \Big)
  \label{eq:AppSigma2} \, .
\end{multline}
These results, Eqs.\ (\ref{eq:AppbA}) and (\ref{eq:AppSigma2}), correspond to Eqs. (\ref{eq:G}) and (\ref{eq:Sigma}).

\section{Analytical results}
\label{app:Anal}

In the low-temperature limit of the Mott insulator with integer filling $n$, the NCA relations can be simplified by observing that the lesser pseudoparticle Green's function $\Gp^<$, corresponding to the occupied pseudoparticle density of states, approaches a delta function in real frequencies
\begin{equation}
  \Gp^<(\omega) \approx
  -i 2 \pi | n \rangle \delta(\omega) \langle n |
  \, . \label{eq:AppLowTApprox}
\end{equation}
This follows from Ref.\ \onlinecite{Kroha:1998aa}, where it is shown that the energies of the projected pseudoparticles are strictly positive and the lesser Green's function is given by $\Gp^<(\omega) = -i2 f_G(\omega) \textrm{Im}[ \Gp^>(\omega) ]$, where $f_G(\omega) = e^{-\beta \omega}$ is the classical Gibbs distribution function. This suppresses all but the local atomic ground state in the zero-temperature limit.
Insertion into the bubble diagram for the spectral function [Eq.\ (\ref{eq:AppbA})] then reduces this expression to
\begin{equation}
  -i 2\pi A(\omega) \approx (n+1) \Gp^>_{n+1}(\omega) - n \Gp^>_{n-1}(-\omega)
  \, .
  \label{eq:AppAapprox}
\end{equation}
So, in the low-temperature limit, the single-particle excitations are the $n-1$ and $n+1$ pseudoparticles with local interaction energies $E_{n \pm 1} - E_n = U/2$.
Also the occupied and unoccupied single-particle states can be approximated according to
\begin{align}
  -i2\pi f(\omega) A (\omega) & \approx n \Gp_{n-1}^>(-\omega) \, , \\
  -i2\pi [ 1 + f(\omega) ] A (\omega) & \approx (n+1) \Gp^>_{n+1}(\omega)
  \, .
\end{align}
Insertion into Eq.\ (\ref{eq:AppSigma2}) gives the simplified greater pseudoparticle self-energy
\begin{multline}
  \Sp^>_m(\omega) \approx
  \frac{i J^2}{2\pi}
  \int_{-\infty}^\infty d\epsilon \,
  \Big[
  n (m+1) \Gp^>_{n-1}(\omega - \epsilon) \Gp^>_{m+1}(\epsilon)
  \\ +
  (n+1) m \, \Gp^>_{n+1}(\omega - \epsilon) \Gp^>_{m-1}(\epsilon)
  \Big]  
  \, ,
\end{multline}
which corresponds to Eq.\ (\ref{eq:SigmaGtrM}).
In particular the $n\pm 1$ pseudoparticles obey the relations
\begin{multline}
  \Sp^>_{n-1}(\omega) \approx
  \frac{i J^2}{2\pi}
  \int_{-\infty}^\infty d\epsilon \,
  \Big[
  n^2 \Gp^>_{n-1}(\omega - \epsilon) \Gp^>_{n}(\epsilon)
  \\ +
  (n+1)(n-1) \, \Gp^>_{n+1}(\omega - \epsilon) \Gp^>_{n-2}(\epsilon)
  \Big]
  \, ,
\end{multline}
and
\begin{multline}
  \Sp^>_{n+1}(\omega) \approx
  \frac{i J^2}{2\pi}
  \int_{-\infty}^\infty d\epsilon \,
  \Big[
  n (n+2) \Gp^>_{n-1}(\omega - \epsilon) \Gp^>_{n+2}(\epsilon)
  \\ +
  (n+1)^2 \, \Gp^>_{n+1}(\omega - \epsilon) \Gp^>_{n}(\epsilon)
  \Big]
  \, .
  \label{eq:SpPartialAppr}
\end{multline}
This can be simplified using the low-temperature approximation [Eq.\ (\ref{eq:AppLowTApprox})] once more, using $\Gp^>_n(\omega) \approx -i2\pi \delta(\omega)$.
However, we go one step further and also apply the numerically motivated approximation of sharp $n \pm 2$ pseudoparticle resonances, i.e., $\Gp^>_{n \pm 2} (\omega) \approx -i 2\pi \delta(\omega - [E_{n \pm 2} - E_n])$, where $E_{n \pm 2} - E_n = 2U$; see also inset in Fig.\ 1.
The local interaction energy $E_m$ of the occupation number state $|m\ra$ is given by $E_m = Um(m-1)/2 - \mu m$  with fixed $\mu = (2n-1)U/2$.
Both of these approximations inserted into Eq.\ (\ref{eq:SpPartialAppr}) give
\begin{multline}
  \Sp^>_{n-1}(\omega)
  \approx
  J^2 n^2 \Gp^>_{n-1}(\omega) +
  J^2 (n^2-1) \Gp^>_{n+1}(\omega - 2U)
  \, , \\
  \shoveleft{
    \Sp^>_{n+1}(\omega)
    \approx
    J^2 (n+1)^2 \Gp^>_{n+1}(\omega)
  } \\ 
  + J^2 n (n+2) \Gp^>_{n-1}(\omega - 2U) 
  \, .
\end{multline}
The first term in these relations, the reappearance of the pseudoparticle Green's function $\Gp_{n\pm 1}^>$ in its own self-energy $\Sp^>_{n \pm 1}$, is the dominant self-energy effect. It acts similar to the Bethe lattice self-consistency relation, producing semicircular spectral functions for the $n \pm 1$ pseudoparticles centered around $\omega = E_{n\pm1} - E_n = U/2$. Their respective bandwidths are directly obtained from the prefactors as $W_{n-1} = 4nJ$ and $W_{n+1} = 4(n+1)J$, in agreement with the Hubbard-I approximation.
The second term in the self-energy expressions is shifted up in energy by $2U$, hence it is located at $\omega = U/2 + 2U = 5U/2$, and it gives a small correction to the spectral functions.

Note that the relations for the $n - 1$ self-energy only hold if $n \ge 2$. In the special case of unity filling $n=1$ there is no $n-2$ Fock state available, and the holon self-energy simplifies to
\begin{equation}
  \Sp^>_0(\omega) \approx J^2 \Gp^>_0(\omega) \, ,
\end{equation}
while the doublon self-energy takes the form
\begin{equation}
  \Sp^>_2(\omega) \approx 4 J^2 \Gp^>_2(\omega) + J^2 3 \Gp^>_0(\omega - 2U)
  \, .
\end{equation}
These are the simplified pseudoparticle self-energy relations in Eqs.\ (\ref{eq:SigmaGtr0}) and (\ref{eq:SigmaGtr2}).
As the holon self-energy $\Sp^>_0$ only depends on the holon propagator $\Gp^>_0$, it follows immediately that $\Gp^>_0$ is semicircular and centered at $\omega = U/2$. By Eq.\ (\ref{eq:AppAapprox}),  $-\Gp^>_0(-\omega)$ directly gives the lower Hubbard band in the spectral function $A(\omega)$. The doublon self energy $\Sp^>_2$ has the same type of semicircular generating term $4J^2 \Gp^>_2$ but also a high-energy triplon plus holon correction emerging in terms of the holon propagator $J^2 3 \Gp^>_0(\omega - 2U)$. Hence in the positive frequency spectral function, the upper Hubbard band is generated by the doublon-doublon self-consistency and the higher triplon excitation by the holon correction term; see Fig.\ 1.

\bibliography{/Users/hugstr/Documents/Papers/DMFT_Biblography}

\end{document}